\documentclass[lettersize,journal]{IEEEtran}
\usepackage{amsmath,amsfonts}
\usepackage{algorithmic}
\usepackage{algorithm}
\usepackage{array}
\usepackage[caption=false,font=normalsize,labelfont=sf,textfont=sf]{subfig}
\usepackage{textcomp}
\usepackage{stfloats}
\usepackage{url}
\usepackage{verbatim}
\usepackage{graphicx}
\graphicspath{ {./} }
\usepackage{cite}
\usepackage{booktabs}
\usepackage{textcomp}

\begin{document}

\title{Simulated Eyeblink Artifact Removal with ICA: Effect of Measurement Uncertainty}

\author{Jennie Couchman, Orestis Kaparounakis, Chatura Samarakoon, and Phillip Stanley-Marbell}

\maketitle

\begin{abstract}
Independent Component Analysis (ICA) is commonly-used in electroencephalogram (EEG) signal processing to remove non-cerebral artifacts from cerebral data. 
Despite the ubiquity of ICA, the effect of measurement uncertainty on the artifact removal process has not been thoroughly investigated.  
We first characterize the measurement uncertainty distribution of a common ADC and show that it quantitatively conforms to a Gaussian distribution. 
We then evaluate the effect of measurement uncertainty on the artifact identification process through several computer simulations. 
These computer simulations evaluate the performance of two different ICA algorithms, FastICA and Infomax,
in removing eyeblink artifacts from five different electrode configurations 
with varying levels of measurement uncertainty. 
FastICA and Infomax show similar performance in identifying the eyeblink artifacts for a given uncertainty level and electrode configuration.
We quantify the correlation performance degradation with respect to SNR and show that in general, an SNR of greater than 15\,dB results in less than a 5\% degradation in performance.
The biggest difference in performance between the two algorithms is in their execution time. 
FastICA's execution time is dependent on the amount of measurement uncertainty, with a 50\% to 85\% reduction in execution time over an SNR range of 20\,dB. 
This contrasts with Infomax's execution time, which is unaffected by measurement uncertainty.
\end{abstract}

\begin{IEEEkeywords}
Independent component analysis, measurement uncertainty, EEG, BCI
\end{IEEEkeywords}

\section{Introduction}
\IEEEPARstart{I}{ndependent} Component Analysis (ICA) is a blind source separation (BSS) technique
that can be used to separate mixtures of independent, non-Gaussian signals.  
ICA has been applied to electroencephalogram (EEG) signal
processing since the late 1990s~\cite{Jung1997} and remains the most
commonly used method to remove non-cerebral artifacts from cerebral
data~\cite{Huster2018, Janani2020,Gu2021b}.  These artifacts arise from
several physical phenomena including eyeblinks, eye movements, muscle movements, and line noise. 
Artifacts present several challenges to researchers and medical professionals, as they generally
have a much higher amplitude than the cerebral signals of interest.
Ocular artifacts such as eyeblinks and eye movements often contaminate the signals measured from the front most EEG electrodes.

Let $\mathbf{s}$ be a random vector of unknown source signals,  $\mathbf{A}$ a linear mixing matrix, and $\mathbf{x}$ a vector of observations of the mixtures of the original source signals. 
Then, we can define a simple mathematical model for ICA as
\begin{equation}
\label{eq:ICA_Def}
\mathbf{x} = \mathbf{A}\mathbf{s}.
\end{equation}
The goal of ICA is to find an unmixing matrix, or a weight matrix,  $\mathbf{W}$, where $\mathbf{W} = \mathbf{A}^{-1}$, such that we can reconstruct the original source signals as 
\begin{equation}
\mathbf{s} = \mathbf{W}\mathbf{x} .
\end{equation}
EEG measurements are widely understood to be weighted linear mixtures of 
underlying electrical biological signals which could be cerebral or artifactual in origin~\cite{Shoker2005,Onton2006, Delorme2007,Makeig2012}.
By applying ICA to recorded EEG data, researchers can identify these underlying signal sources.
Then, they can remove the artifactual signals and re-mix the data to create a clean EEG recording that is not contaminated by unwanted signals.

Despite the ubiquity of ICA in EEG signal processing, the effect of measurement
uncertainty on the output of ICA has not been thoroughly analyzed. 
In fact, most ICA algorithms are designed assuming that the observed signals are noiseless~\cite{James2005}.
The performance of any signal processing algorithm is generally a function of the Signal-to-Noise Ratio (SNR) of the input variables. 
Here, we consider the \emph{signal} of an EEG measurement to be any biological electrical signal, whether it is sourced in the brain or another biological process.
We consider the \emph{noise} to be any non-biological component in the measurement.
This noise can arise from several physical phenomena, including inaccuracies and uncertainties in the measurement environment and measurement tools.
For this reason, we consider measurement uncertainty to be a large portion of the noise present in the signal.
With this in mind, the output of ICA is expected to change based on the amount of measurement uncertainty.  
It is important for analysts to understand the ways in which the results of ICA are affected by measurement uncertainty.  
If there is a large amount of uncertainty in the measurement system, then the results of ICA may be unreliable and further verification of the outputs may be needed. 
Conversely, if the amount of uncertainty in the system is sufficiently low, analysts can be confident that the measurement uncertainty had a negligible effect on the results of ICA.

Through the last several decades, many ICA algorithms have been proposed~\cite{Comon1994, Hyvarinen1999a,Bell1995b, Lee1999,
Learned-Miller2003,Koldovsky2005,Chen2015a}.
The two ICA algorithms most commonly used for EEG artifact removal are 
Hyv{\"{a}}rinen's fixed-point algorithm FastICA~\cite{Hyvarinen1999a} and
Bell and Sejnowski's algorithm Infomax ICA~\cite{Bell1995b}.
We choose to focus on these two algorithms in this analysis because
they are both available through two commonly-utilized open-source EEG toolboxes: 
EEGLAB~\cite{Delorme2004} for MATLAB and MNE~\cite{Gramfort2013} for Python.

In this paper we make the following contributions:

\begin{enumerate}
    \item Validation of the commonly-held assumption that measurement
    uncertainty in EEG data has a Gaussian distribution. 
    \item Demonstration of the relationship between the amount of measurement
    uncertainty in EEG data and the ability of ICA to identify an eyeblink
    artifact for FastICA and Infomax ICA. 
    \item Insights on the relationship between measurement uncertainty and the
    execution time for both FastICA and Infomax ICA.
    \item Experimental quantification of the degradation of eyeblink
    identification performance of ICA with respect to SNR. This information
    enables researchers and designers to know in advance how measurement
    uncertainty affects the expected efficacy of ICA.
\end{enumerate}

\section{Experimental Analysis}
\label{sec:Experiments}
\subsection{Experimental Setup}
To characterize measurement uncertainty in an EEG system, we evaluate the uncertainty present in the
OpenBCI Cyton Board~\cite{OpenBCI2022}. The Cyton Board uses the Texas
Instruments ADS1299~\cite{TexasInstruments2017}, an analog-to-digital converter (ADC)
that is designed to work with biopotential measurements. 
In 2018, Rashid \textit{et al.}~\cite{Rashid2018} found that despite its relatively low cost, 
systems built with the ADS1299 showed comparable performance to high-end medical grade
EEG systems, so we consider the ADS1299 to be an acceptable representation of commonly-used EEG hardware.
Per the manufacturer's specification the ADS1299 has a low input-referred noise of $0.1$\,\textmu V peak-to-peak.
The maximum allowed analog input voltage to the ADS1299 is the supply voltage plus $0.3$\,V. 
The Cyton board has an analog voltage regulation of $2.5$\,V, so the maximum
voltage allowed input signal amplitude is $2.8$\,V.  When used in an 8-input
configuration, the Cyton board uses a sampling frequency of 250\,Hz. 

\begin{figure}[!t]
\centering
\subfloat[]{\includegraphics[width=2.5in]{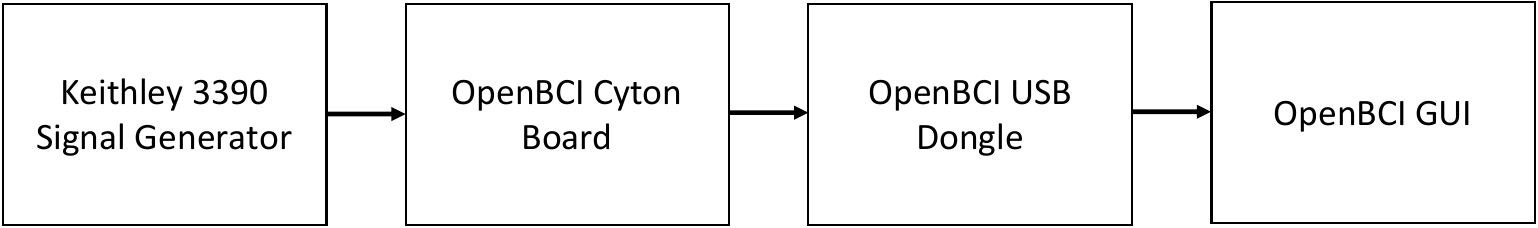}%
\label{fig:experimentDiagram}}
\vfil
\subfloat[]{\includegraphics[width=2.5in]{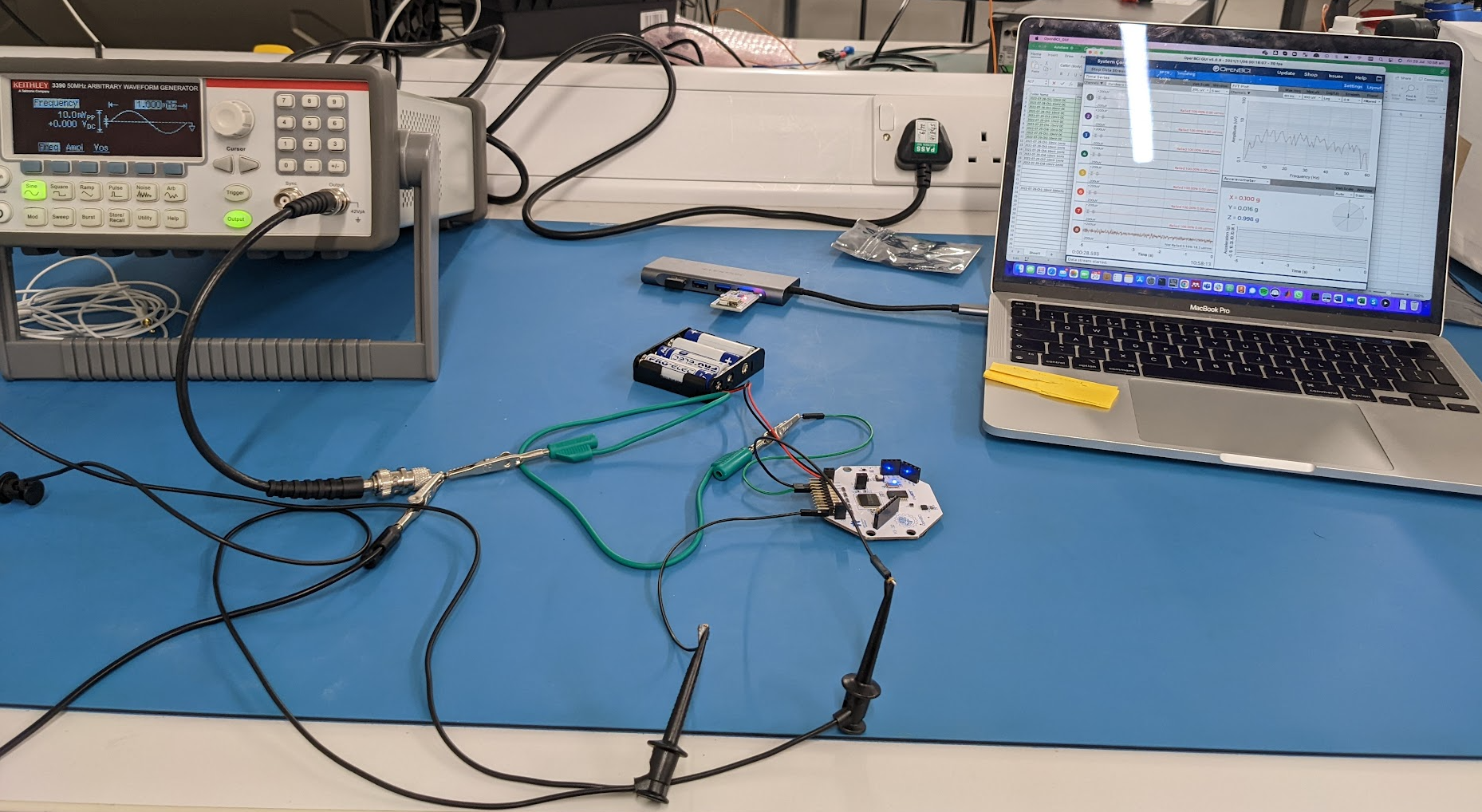}
\label{fig:experimentPhoto}}
\caption{The experimental setup for our analysis.The Keithley 3390 arbitrary waveform generator is directly coupled to the input pins of the OpenBCI Cyton Board 
through the use of crocodile clips. The data is transmitted via an RF connection to a USB dongle plugged into the computer. 
The OpenBCI GUI allows the user to view the data measurement in real time and save off data recordings.}
\label{fig:experiment}
\end{figure}

Figure \ref{fig:experiment} shows a diagram and a picture of the experimental setup we use for our analysis. We use
a Keithley 3390 arbitrary waveform generator to generate an input waveform with a
frequency of $1$\,\textmu Hz,  the lowest possible output frequency for this generator. 
We directly couple the output of the waveform generator to different measurement pins of the Cyton board
from recording to recording and use a peak-to-peak amplitude of 10\,mV.   
We record the data using the OpenBCI GUI and save each of the recordings in a separate CSV file.  
The recordings had an average of 18,000 data points each.

\subsection{Uncertainty Characterization and Analysis}
\label{section:UncertaintyCharacterization}

To keep analysis consistent for each of the recordings and to eliminate transient effects of starting and
stopping the data recordings, we truncate each recording to keep the middle 60 seconds. This corresponds to 15,000 data points. 
For a sinusoidal signal frequency of $1$\,\textmu Hz and a peak-to-peak voltage of $10$\,mV, the maximum variation of the signal in this 60 second range is $1.9$\,\textmu V.  
Such low variation in the signal amplitude over this time period means that the
true value of underlying measurand can be considered constant in our analysis.
As such, we are not considering them as a time-varying series. Instead, we
consider them as repeated measurements of the same underlying value.

\begin{figure}[!t]
\centering
\includegraphics[width=2.5in]{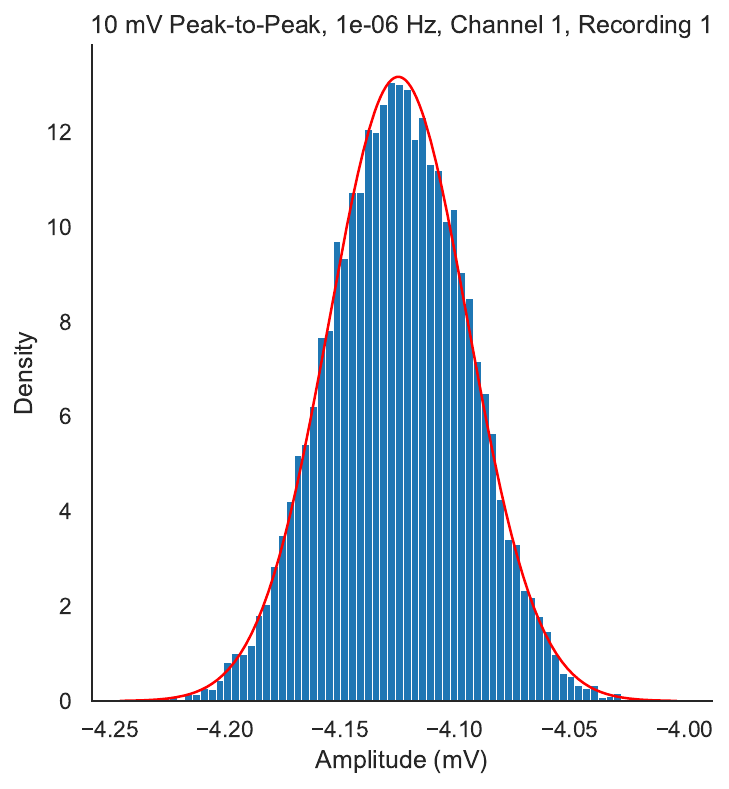}%
\caption{The distribution of a measurement recording of an approximately constant signal. Each of the histogram bins has a width of approximately $3.4$\,\textmu V. While the voltage of the true signal can be considered constant for the data recording, the measurement distribution has an apparent Gaussian shape. }
\label{fig:10mV_Recording1}
\end{figure}

\begin{figure}[!t]
\centering
\includegraphics[width=2.95in]{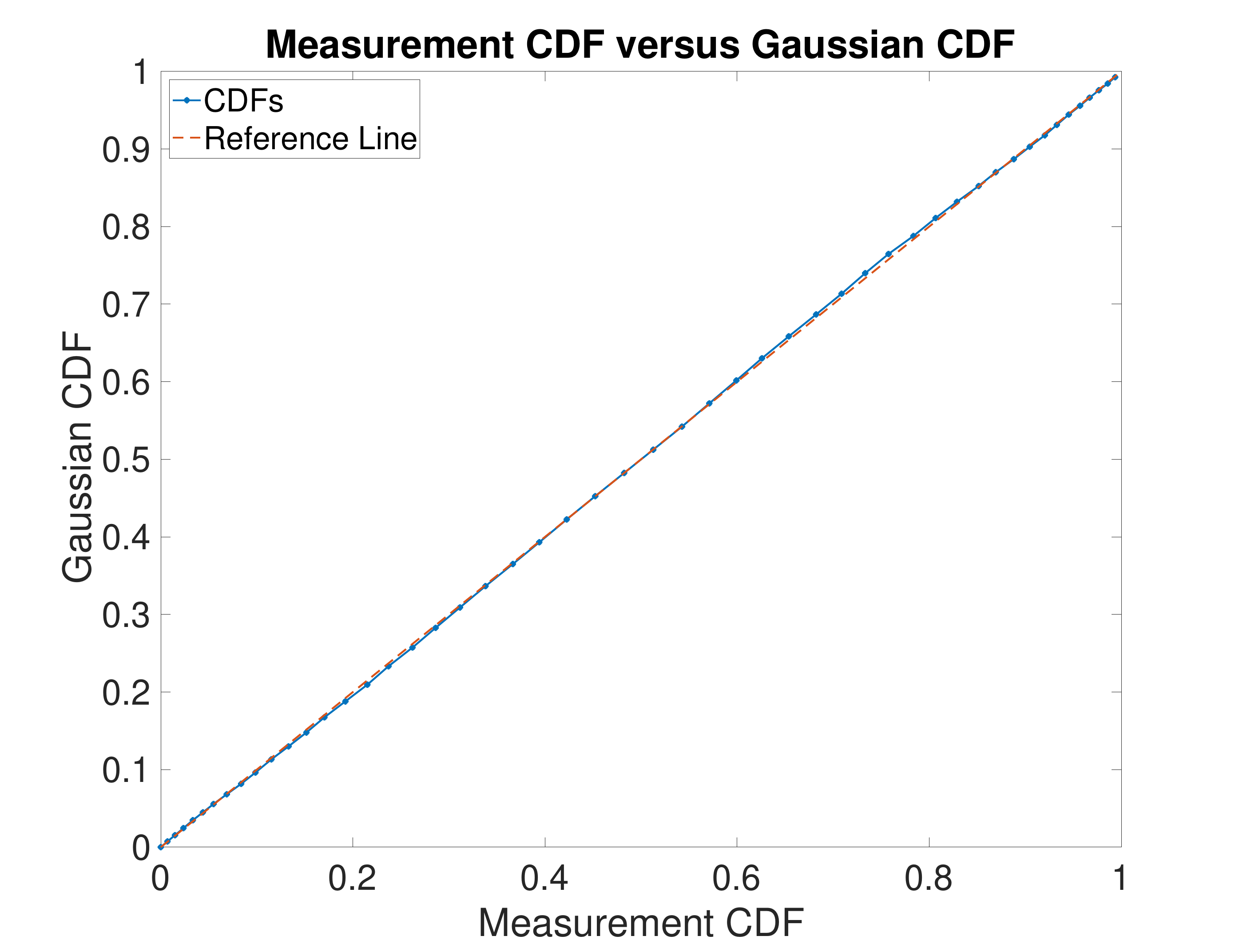}
\caption{A comparison between the measurement record's CDF and the CDF of a Gaussian with the same mean and variance. The CDFs are almost exactly aligned, furthering the evidence that we can consider the measurement uncertainty to be Gaussian in nature.}
\label{fig:cdf_comparison}
\end{figure}

Figure \ref{fig:10mV_Recording1} shows the histogram of one of our measurement recordings.  
As it is a histogram of repeated measurements of the same underlying value, the histogram represents the uncertainty distribution of the measurement.  
The width of each histogram bin is around $3.4$\,\textmu V, approximately $1.8$ times greater than the maximum voltage swing of $1.9$\,\textmu V. 
The distribution has an apparent Gaussian shape. 
Figure \ref{fig:cdf_comparison} shows a comparison of the measurement's CDF and the CDF of a Gaussian distribution with the same mean and variance. 
The CDFs are lined up almost exactly, furthering the evidence that this record has a Gaussian shape.
The skewness of this distribution is $0.0016$ and the kurtosis of this distribution is $2.92$.
These two properties are close to the ideal Gaussian skewness of $0$ and kurtosis of $3$.
This measurement record is representative of our results. 
Based on these quantitative and qualitiative measures, we use Guassian distributions to model measurement uncertainty in Section \ref{sec:Simulations}. 
The expected amplitudes in a real EEG recording tend to be on the order of hundreds of microvolts,
so the $10$\,mV peak-to-peak amplitude we use as the input to our measurement recordings are not representative of EEG data.
For this reason, we do not use the experimentally-obtained uncertainty from this section in any of our further sections. 
Instead, we use synthetically generated Gaussian noise of appropriate amplitudes.

\section{Synthetic Data Generation}
\label{sec:DataGeneration}
It is difficult to use real EEG data to evaluate algorithms due to a lack of
knowledge about the \emph{ground truth} of the signal. We instead use synthetic
data for our analysis.  This allows us to more accurately evaluate the ability
of ICA to properly separate artifacts from the rest of the EEG data.  
We started with real EEG data originally collected by Tanner~\cite{Tanner2019}
in 2019. This data was anonymized after collection and made open source. We
retrieved the raw data from Harvard Dataverse~\cite{Tanner2019Data}. Tanner
collected the data using the Brain Vision BrainAmp DC~\cite{BrainVision}. 
The data files were recorded by applying an analog low-pass anti-aliasing filter with a cutoff frequency of 250\,Hz to each signal, 
followed by sampling the signals with a sampling frequency of 1000\,Hz.
The data have an amplitude resolution of 0.1\,\textmu V.  
This corresponds to a quantization noise level of approximately 0.0289\,\textmu V RMS. 

\begin{figure}[!t]
\centering
\includegraphics[width=2.5in]{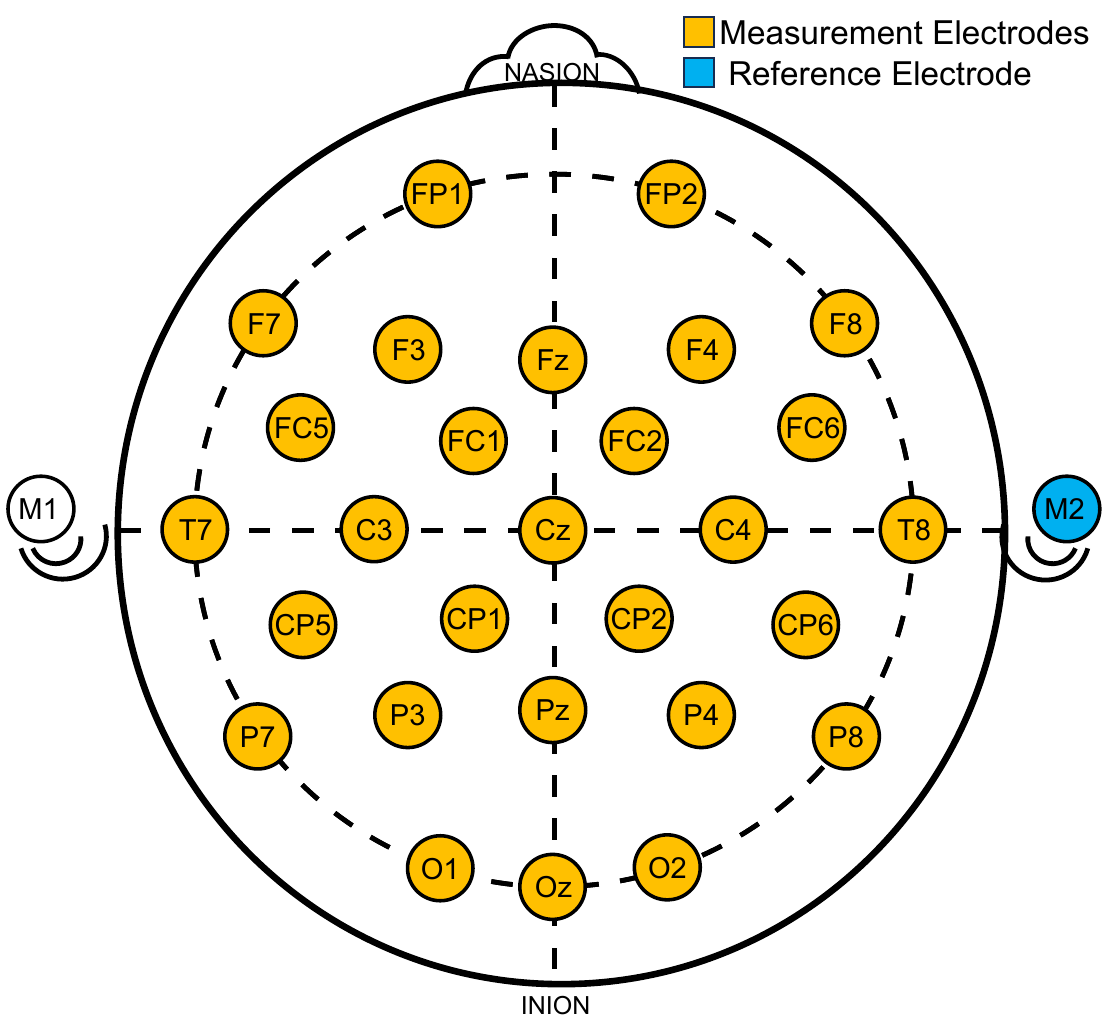}
\caption{The electrode locations of the EEG channels present in the data. The orange electrodes were considered measurement electrodes, and the blue electrode (M2) was used as the reference electrode.}
\label{fig:ElectrodeMap}
\end{figure}

The data recordings contained 28 EEG measurement channels, 1 EEG reference channel, and 3 electrooculography (EOG) measurement channels for each subject. 
EOG signals are produced by eye movements and are recorded by electrodes that are placed near the eyes.
In these data sets, the EOG signals are used as reference channels that aid in identifying ocular artifacts due to eyeblinks and eye movements.  
The EEG channel names refer to their placement on the head according to the International 10-20 standard~\cite{Chatrian1985} for EEG measurement.
The EOG channels are placed at the LO1, LO2, and IO1 positions.
These positions correspond to the outside corner of the left eye, the outside corner of the right eye, and underneath the left eye, respectively.
Figure \ref{fig:ElectrodeMap} shows the locations of all the EEG measurement and reference electrodes present in the data.

\subsection{Dataset Synthesis Pipeline}
We use real raw EEG data from six subjects~\cite{Tanner2019Data}. We synthesize a
clean eyeblink independent component segment which we apply to all subjects
after pre-processing.

After performing ICA\footnote{Python MNE FastICA with random state 42.} on the
first subject, 101, we compute the Pearson correlation between each independent
component and the filtered EOG channels. Using adaptive z-scoring and a 2.6
threshold we isolate two components for eye-related artifacts. We select the
eyeblink component (ICA002) after visually inspecting the components and
comparing with insights about eyeblink and eye movement components in
literature~\cite{Chaumon2015}. We crop an one-second segment of one eyeblink
and we low-pass filter it with a cut-off frequency of 35\,$\mathrm{Hz}$. We
nullify the other components and un-mix the solution back to the original
channels. In this way we have the effect of the one-second eyeblink segment on
each of the source channels. 

We pre-process each of the subject datasets to remove low-frequency drift and
eyeblink artifacts to create a clean dataset. We high-pass filter the channel
data with a cut-off frequency of 1\,$\mathrm{Hz}$. With the same configuration
as for isolating the eyeblink related components on subject 101, we perform ICA
on each subject to nullify the isolated components before un-mixing the data
again. This creates a clean dataset that also has no eyeblink artifacts.

To create a simulated dataset for each subject, we add the one-second-long
eyeblink artifact segment to random timestamps of the cleaned subject. Each
random timestamp value follows a uniform distribution with a five-second interval
after a five-second offset from the previous timestamp, i.e., $T_i \sim
\mathrm{U}(t_{i-1}+5, t_{i-1}+10)$, where $t_{i-1}$ is the sampled value for the
previous timestamp and $t_0 = 0\,\mathrm{s}$.

To create a synthetic reference channel for the simulated eyeblink we set
channel LO1 equal to the simulated eyeblink independent component. The
simulations in Section~\ref{sec:SimulationsSimulations} use LO1 as a reference
channel.

\section{Independent Component Analysis Simulations with Varying SNR}
\label{sec:Simulations}
We evaluate the effect of varying amounts of measurement uncertainty on the artifact separation performance of ICA when applied to EEG data. 
All signal processing algorithms are dependent on the Signal-to-Noise Ratio (SNR) of the input data. 
Here, we consider measurement uncertainty to be part of the ``noise'' and any physiological measurement to be the true signal. 

\subsection{EEG Electrode Configurations}
\label{sec:configs}
\begin{figure*}[!t]
\centering
\subfloat[com9]{\includegraphics[width=2.5in]{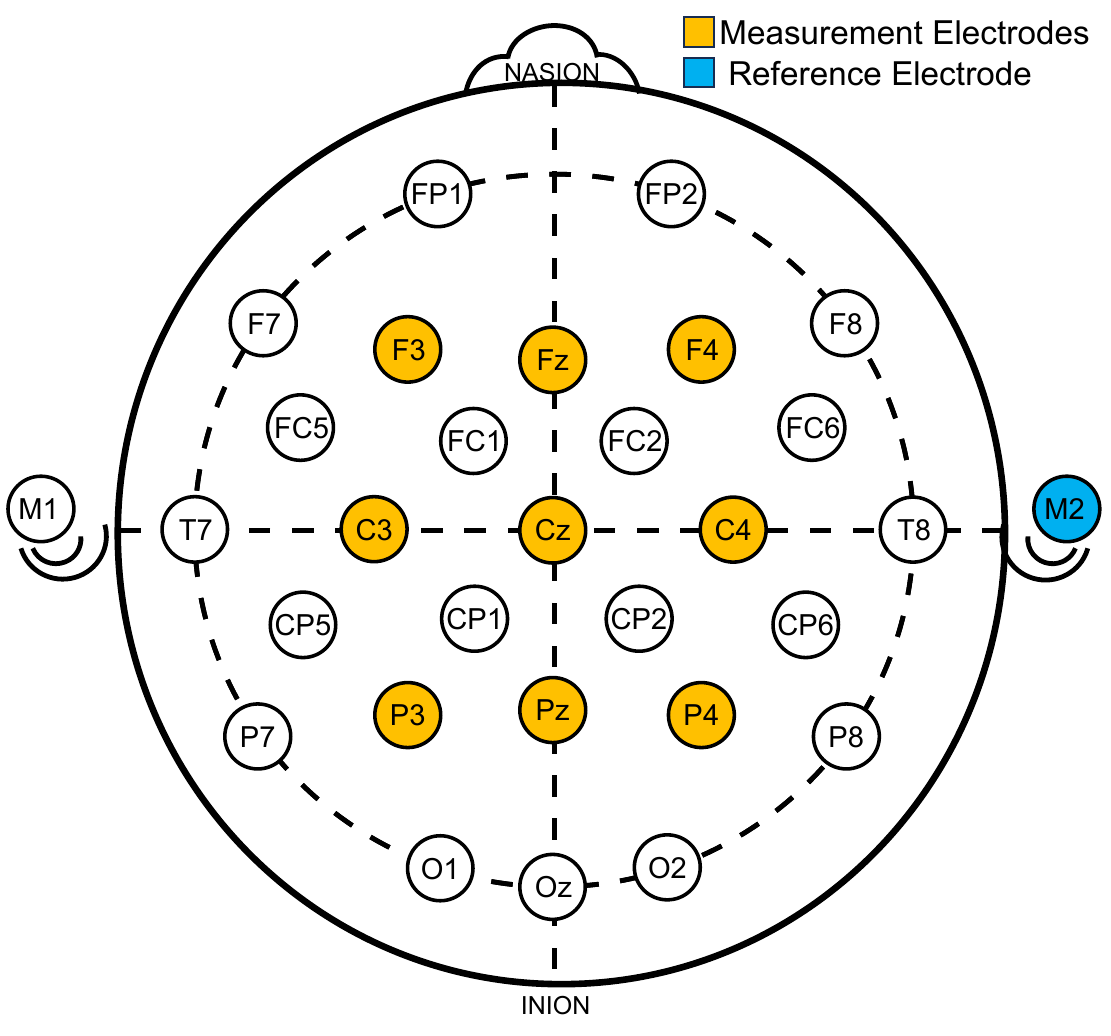}%
\label{fig:com9locations}}
\hfil
\subfloat[em8]{\includegraphics[width=2.5in]{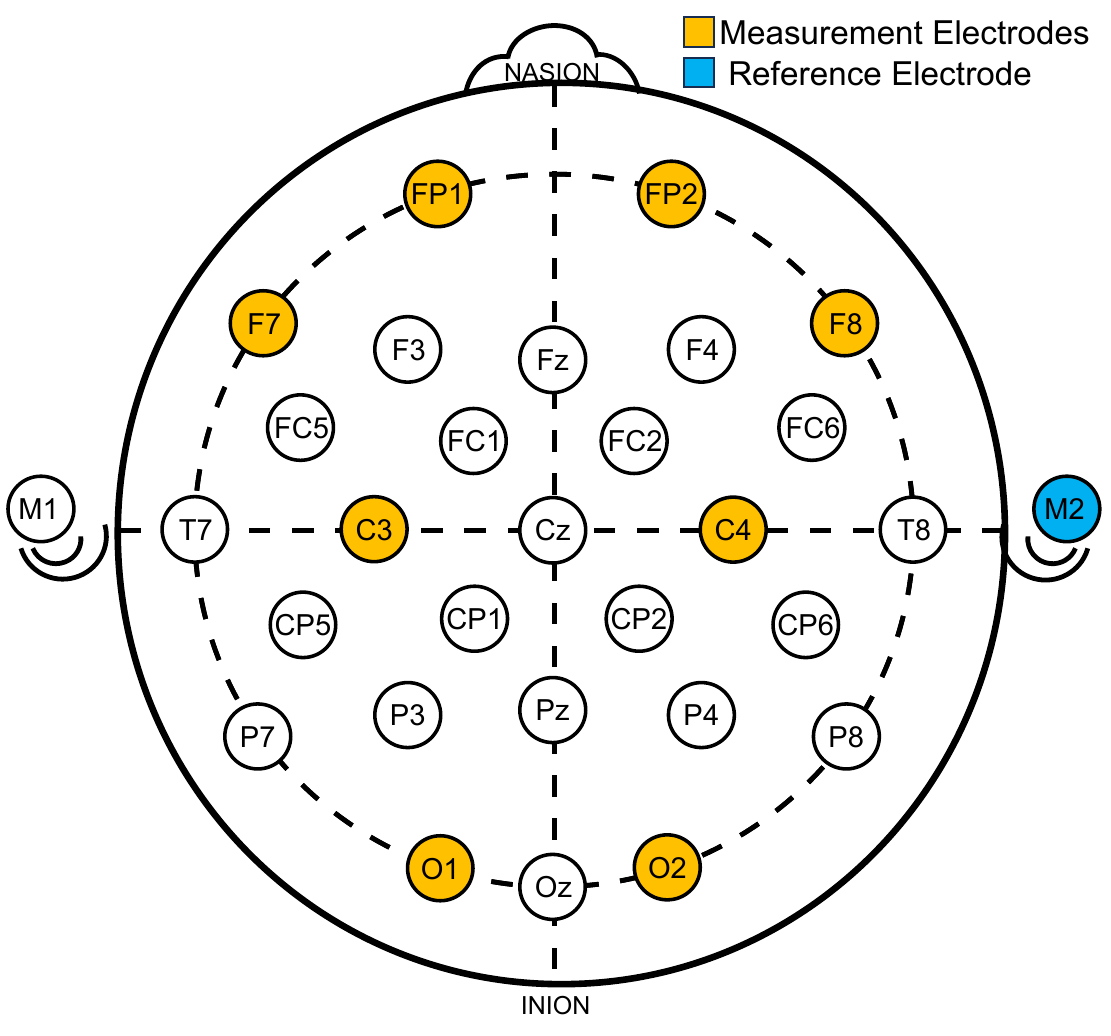}%
\label{fig:em8locations}}
\vfil
\subfloat[att8]{\includegraphics[width=2.5in]{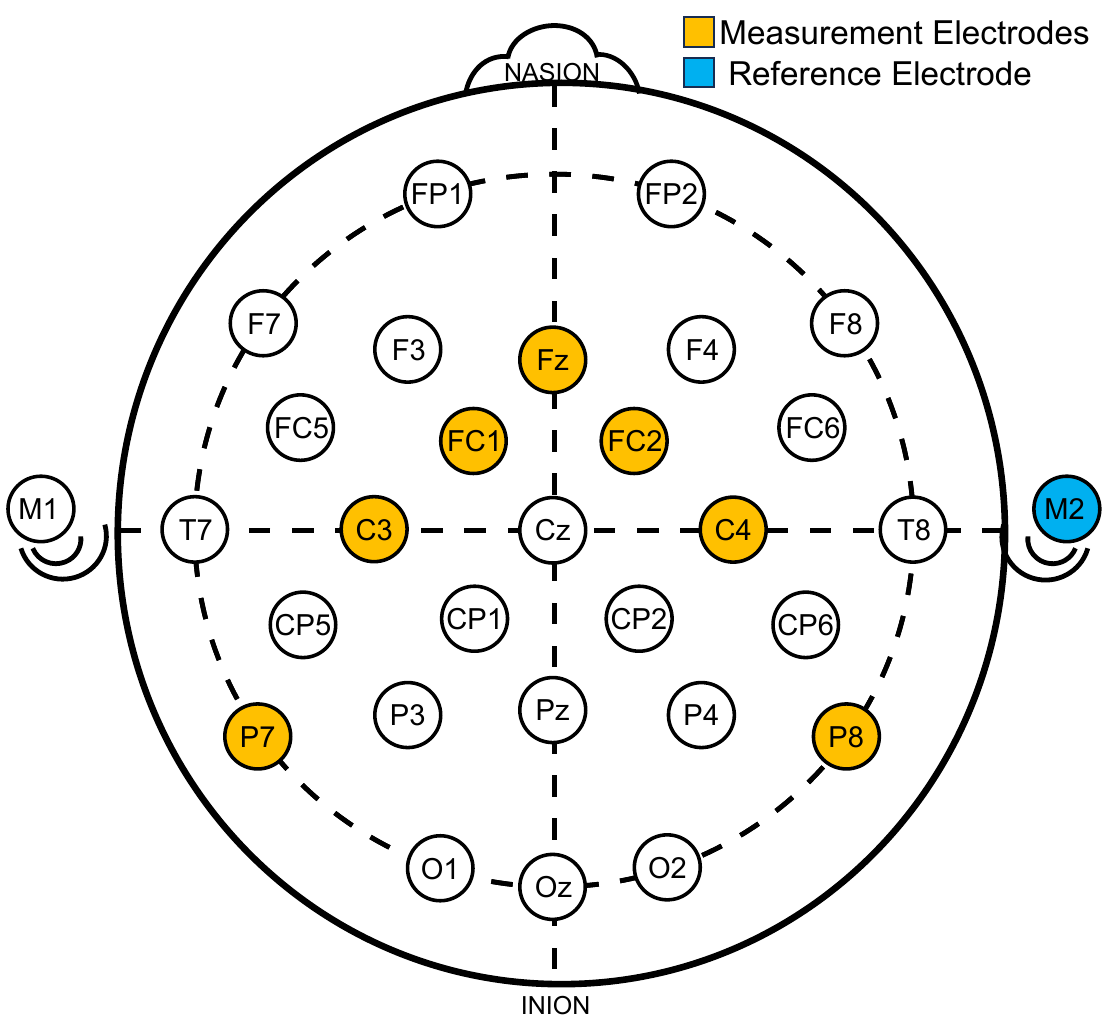}
\label{fig:att8locations}}
\hfil 
\subfloat[mi10]{\includegraphics[width=2.5in]{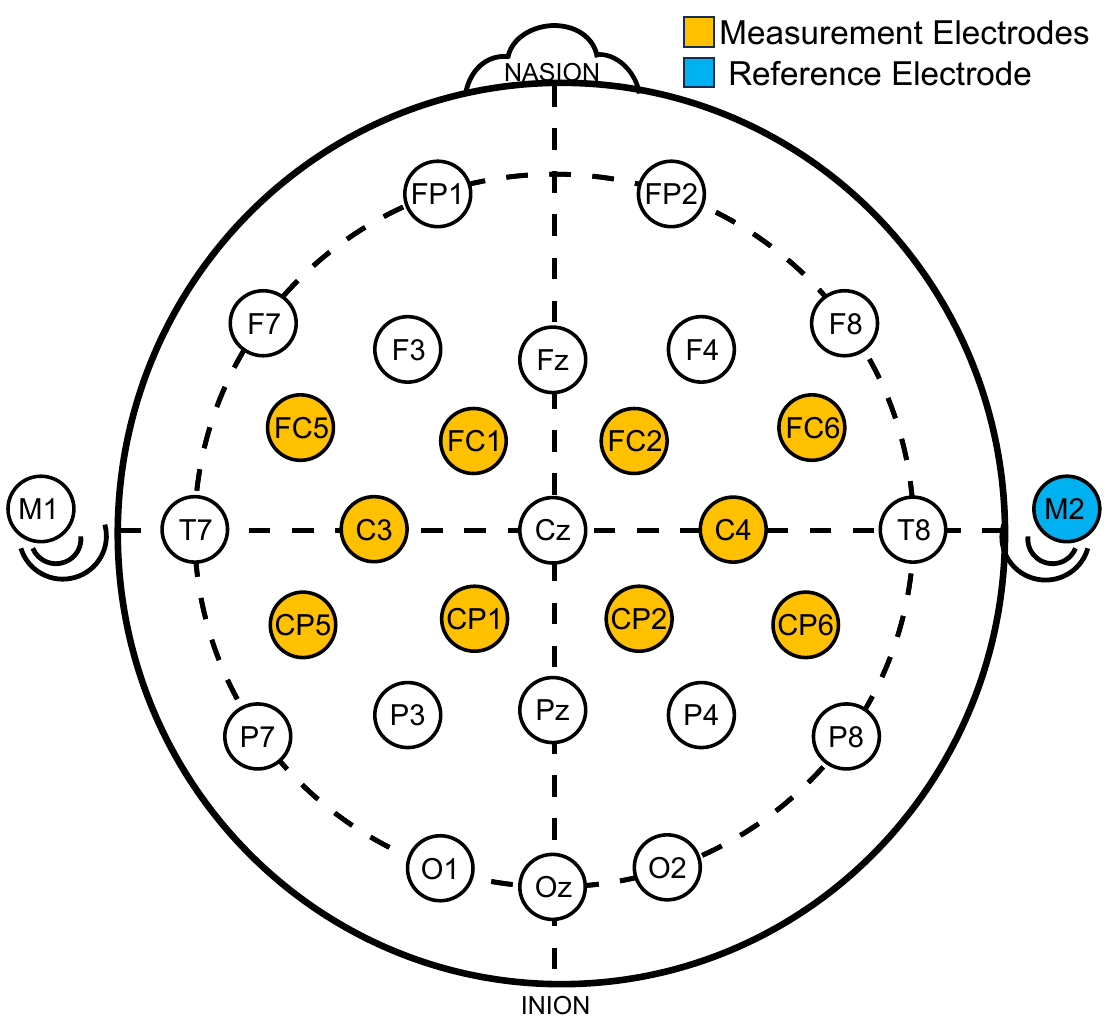}}
\label{fig:mi10locations}
\caption{The four additional electrode configurations used in our simulations.  The orange electrodes are measurement electrodes and the blue electrode is a reference electrode. The white electrodes are unused in a particular configuration. (a) Electrodes corresponding to the Biopac Systems B-Alert X10 BCI,  the configuration we refer to as ``com9'' (b) Electrodes selected to optimize for emotion-specialized tasks, the configuration we refer to as ``em8''. (c) Electrodes selected to optimize for attention-specialized tasks, the configuration we refer to as ``att8''. (d) Electrodes selected to optimize for motor imagery tasks, the configuration we refer to as ``mi10''.}
\label{fig:ElectrodeSubsetLocations}
\end{figure*}

The synthetic dataset we generated in Section \ref{sec:DataGeneration} utilizes 28 EEG channels that cover the whole head. 
In practice, EEGs and BCIs are often used in specialized tasks that only require a smaller subset of electrodes. 
We analyze the effect of measurement uncertainty on the ability to remove eyeblink artifacts from EEG measurements 
taken from four specific electrode configurations in addition to the full set of channels.
Figure \ref{fig:ElectrodeSubsetLocations} shows these electrode configurations.
We chose the first configuration of electrodes to correspond the ones used in the commercial Biopac B-Alert X10 BCI headset~\cite{BiopacSystemsInc.2023}.  
We refer to this subset as the ``com9'' configuration.
We selected the next two configurations to closely resemble the 8-electrode configurations proposed by Park \textit{et al.}~\cite{Park2020} for emotion-specialized and attention-specialized tasks. 
We refer to these subsets as the ``em8'' and ``att8'' configurations, respectively.
We chose the final subset to match BCIs used for motor imagery tasks~\cite{Zakrzewski2022}.
We refer to this subset as the ``mi10'' configuration.

\subsection{Simulations}
\label{sec:SimulationsSimulations}
We start with the datasets we generated in Section \ref{sec:DataGeneration} as the inputs to our simulations. 
For each iteration of the simulation,  we select a random one-minute segment of data from the continuous data, which corresponded to $60,000$ contiguous time samples.
Selecting a smaller time segment for each iteration allowed for a speedier simulation time while ensuring the number of samples far exceeds existing recommendations on the minimum window length for ICA given the maximum number of channels in our datasets~\cite{Onton2006a,Korats2013}.  
We then simulate measurement uncertainty on each EEG channel in the form of additive Gaussian noise, to reflect the Gaussian shape of the uncertainty distributions we characterized in Section \ref{section:UncertaintyCharacterization}. 
We add Guassian noise at a specified SNR to each EEG channel.
We do not add any measurement uncertainty to the channel labeled LO1, as this channel is not used in the ICA algorithms and instead is used solely as a reference channel for post-processing and analysis.

We use six (6) synthetic datasets in our simulations. 
For each dataset, we run simulations utilizing all available channels as well as the four channel subsets we describe in Section \ref{sec:configs} and display in Figure \ref{fig:ElectrodeSubsetLocations}.
We run both FastICA and Infomax ICA with varying amounts of measurement uncertainty. 
We accomplish this by adding Gaussian noise to each channel at SNRs ranging from $0$\,dB to $20$\,dB in $2.5$\,dB increments. 
We use the same SNR for every channel in a given simulation run. 
In total we run 540 simulations, each one with 100 re-executions selecting a different random one-minute segment in each execution (akin to doing a Monte Carlo simulation).

\section{Simulation Results}
\label{sec:Results}
\begin{figure*}[!t]
\centering
\subfloat[]{\includegraphics[width=3in]{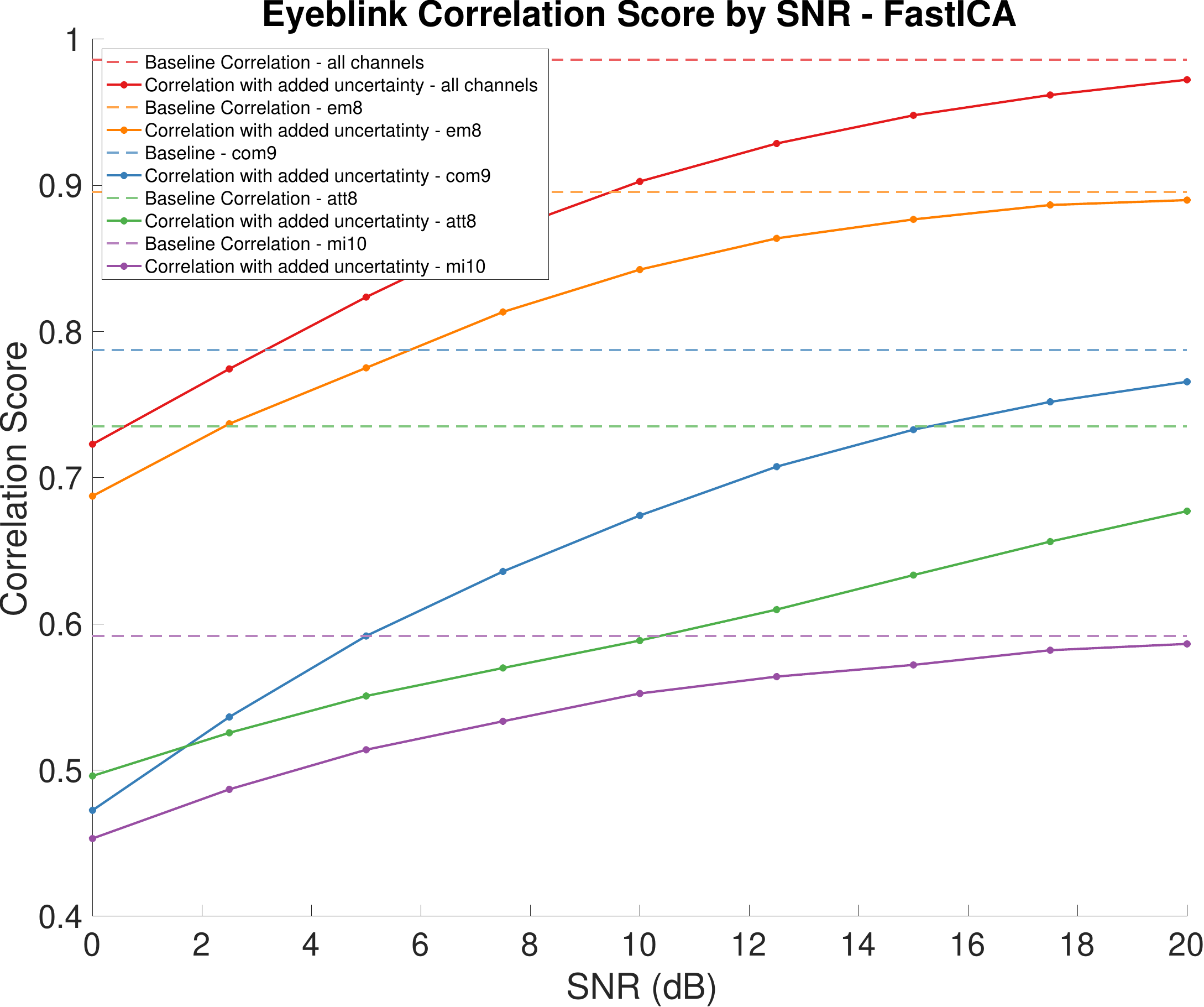}}
\hfil 
\subfloat[]{\includegraphics[width=3in]{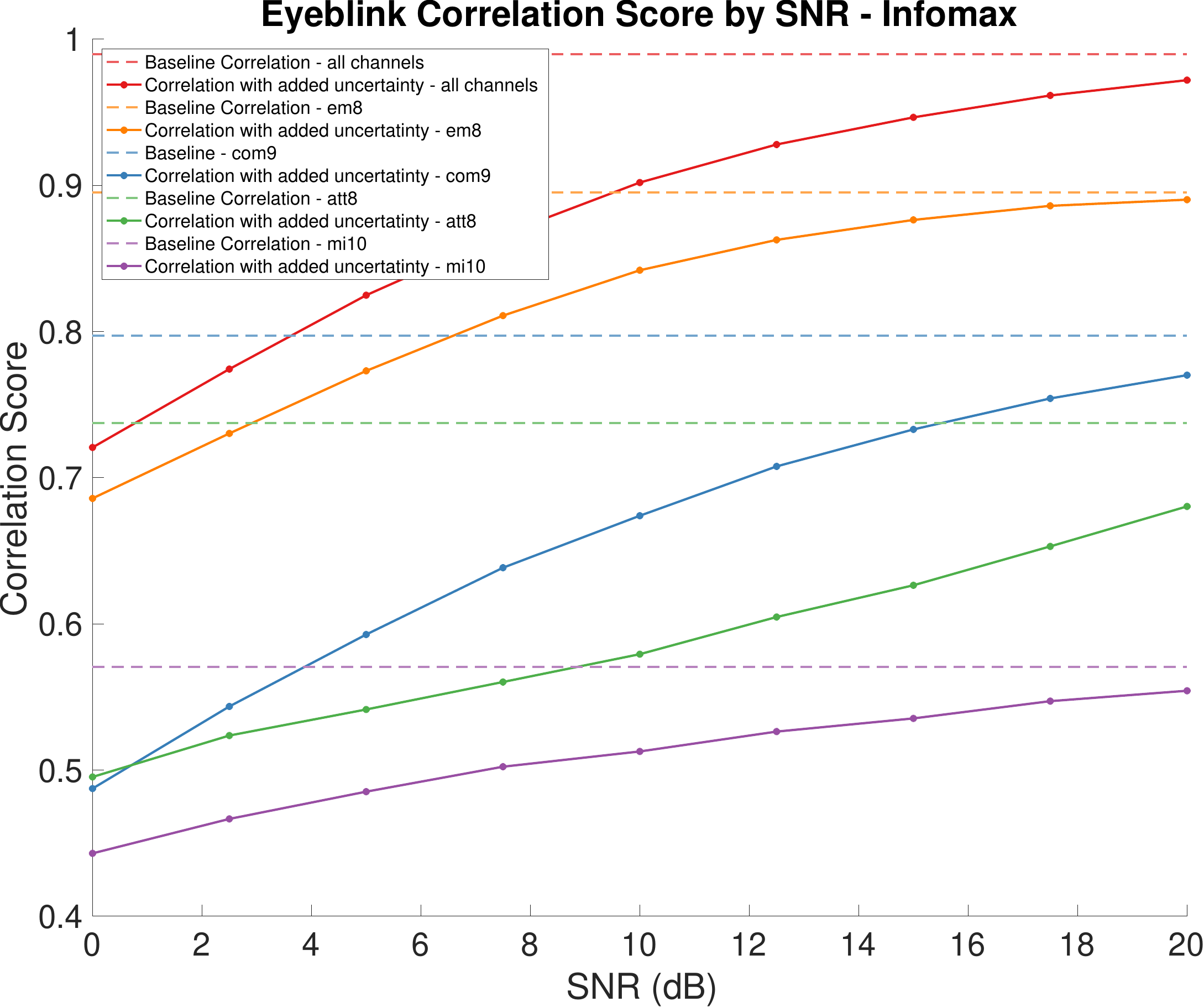}}
\caption{The absolute values of the Pearson correlation scores with the synthetic eyeblink artifacts. FastICA and Infomax show comparable correlation performance with very similar correlation values at each SNR level, particularly for the electrode configurations that include the front-most electrodes. }
\label{fig:CorrelationScores}
\end{figure*}

To determine the artifact separation performance at each iteration, we compute the absolute value of Pearson correlation between every independent component identified and our reference eyeblink signal. 
We define the ``correlation score'' of a given iteration to be the maximum Pearson correlation value between any independent component and the reference eyeblink signal. 
We also compute the \emph{baseline} correlation score by running the simulations with no measurement uncertainty added. 
Figure \ref{fig:CorrelationScores} shows the aggregated results of our simulations, with the FastICA results on the left and the Infomax results on the right. 
The two algorithms produced very similar correlation performance results. 
As expected, the correlation values increase with SNR. 

\begin{figure*}[!t]
\centering
\subfloat[]{\includegraphics[width=1.25in]{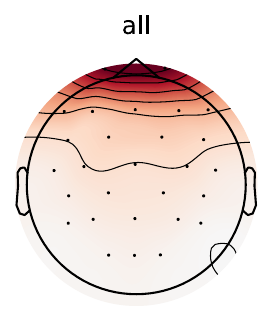}}
\hfil 
\subfloat[]{\includegraphics[width=1.25in]{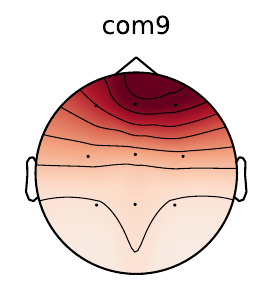}}
\hfil 
\subfloat[]{\includegraphics[width=1.25in]{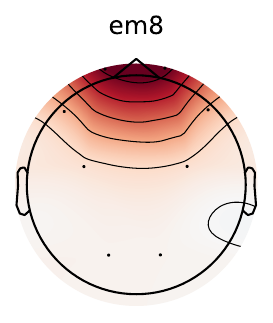}}
\hfil 
\subfloat[]{\includegraphics[width=1.25in]{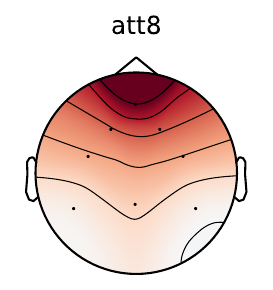}}
\hfil 
\subfloat[]{\includegraphics[width=1.25in]{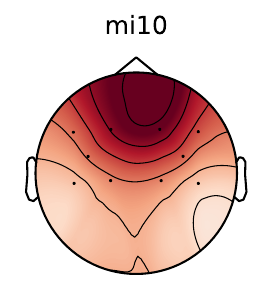}}
\caption{Topography maps of the contribution of each sensor to the independent component most associated with the eyeblink artifact based on the baseline Pearson correlation scores for each electrode configuration.  
Deeper colors represent a higher weight corresponding to the co-located channel.  
The sensors locations are denoted by the dots on all the subfigures (a)-(e). 
The independent component topography maps corresponding to the configurations (a) ``all''  and (c) ``em8''  configuration are very typical of eyeblink artifacts, 
with high weights associated with the electrodes placed at the front of the head and lower weights associated with all the other electrodes.  
The topography maps corresponding to the configurations (b) ``com9'', (d) ``att8'', and (e) ``mi10'' also show strong activity at the front of the head but still show a high amount of activity further back in the head. 
This indicates that these components contain a significant portion of cerebral data in addition to the eyeblink artifact.  
Most experienced researchers would not consider these three to be valid eyeblink components.}
\label{fig:EyeblinkComponents}
\end{figure*}

Correlation scores are highest for electrode configurations that include the front-most electrodes, FP1 and FP2. 
These configurations are the ones where all electrodes are used (``all'') and the emotion-specialized measurements (``em8''). 
This is very intuitive, as these two electrodes are physically closest to the eyes, so measurements taken from these electrodes are likely to receive a large amount of the electrical signal generated by eyeblinks. 
The low correlation scores associated with all of the other configurations indicate that in these cases, ICA would not be suitable to identify and remove eyeblink artifacts from the EEG data.
Figure \ref{fig:EyeblinkComponents} shows the topography maps\footnote{MNE plots the FP1 and FP2 sensor locations slightly outside the head circle on the sensor location maps.
For this reason, the ``all'' and ``em8'' topography maps have contour regions outside the bounds of the head circles.} 
of the independent components most closely corresponding to the reference eyeblink signal based on the Pearson correlation score.
From these topography maps, we see that the components associated with the ``all'' configuration and the ``em8'' configuration are the only ones that can be visibly verified as eyeblink components based on standard practice~\cite{Chaumon2015}.

\begin{figure*}[!t]
\centering
\subfloat[]{\includegraphics[width=3in]{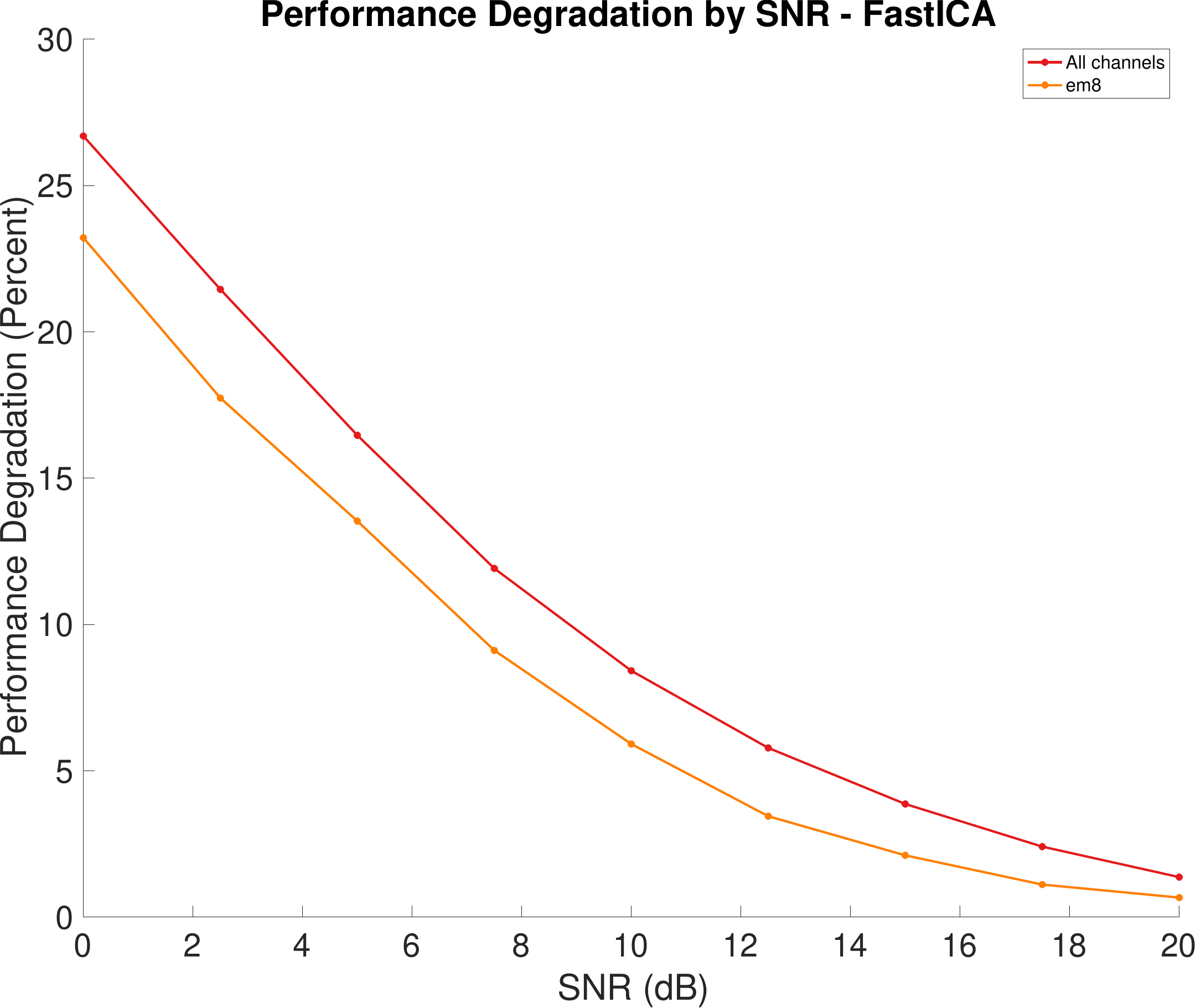}}
\hfil 
\subfloat[]{\includegraphics[width=3in]{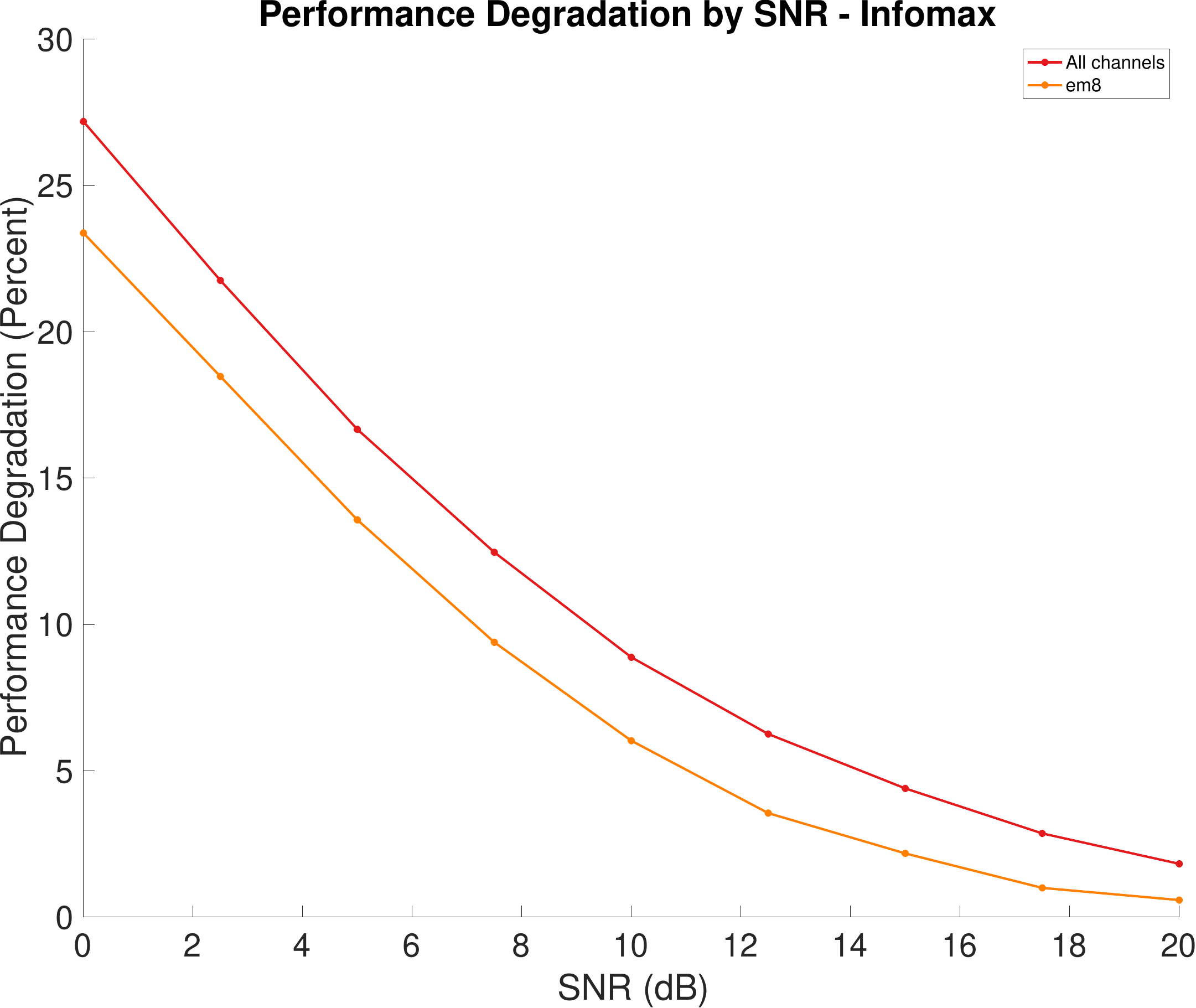}}
\caption{The correlation performance degradation from the baseline scores when measurement uncertainty is added.  In general, an SNR of greater than $15$\,dB corresponds to a performance degradation of less than $5\%$.}
\label{fig:PerformanceDegradation}
\end{figure*}

Because the ``all'' and ``em8'' configurations are the only configurations that find an independent component 
that would be reasonably identified as an eyeblink by experienced clinicians and researchers, 
we now focus on these two configurations and quantify
how much the correlation performance changes with varying amounts of uncertainty.
To make this comparison, we define performance degradation $q$ at a given SNR $\gamma$ as
\begin{equation}
q_{\gamma} = \frac{\rho_{0}-\rho_{\gamma}}{\rho_{0}}
\end{equation}
where $\rho_{0}$ is the baseline correlation score for a given electrode configuration and $\rho_{\gamma}$ is the correlation score for the same electrode configuration with SNR $\gamma$.
Figure \ref{fig:PerformanceDegradation} shows this performance degradation for the electrode configurations ``all'' and ``em8'' with FastICA on the left and Infomax on the right. 
For both configurations an SNR of approximately $7.5$\,dB corresponds to an approximate $10\%$ performance degradation and an SNR of $15$\,dB corresponds to an approximate $5\%$ performance degradation.
An SNR of $7.5$\,dB means that the biological signal's RMS amplitude is around $2.4$ times higher than the RMS amplitude of the uncertainty.
Likewise, an SNR of $15$\,dB indicates that the biological signal's RMS amplitude is around $5.6$ times higher than the RMS amplitude of the uncertainty.

\begin{figure*}[!t]
\centering
\subfloat[]{\includegraphics[width=3in]{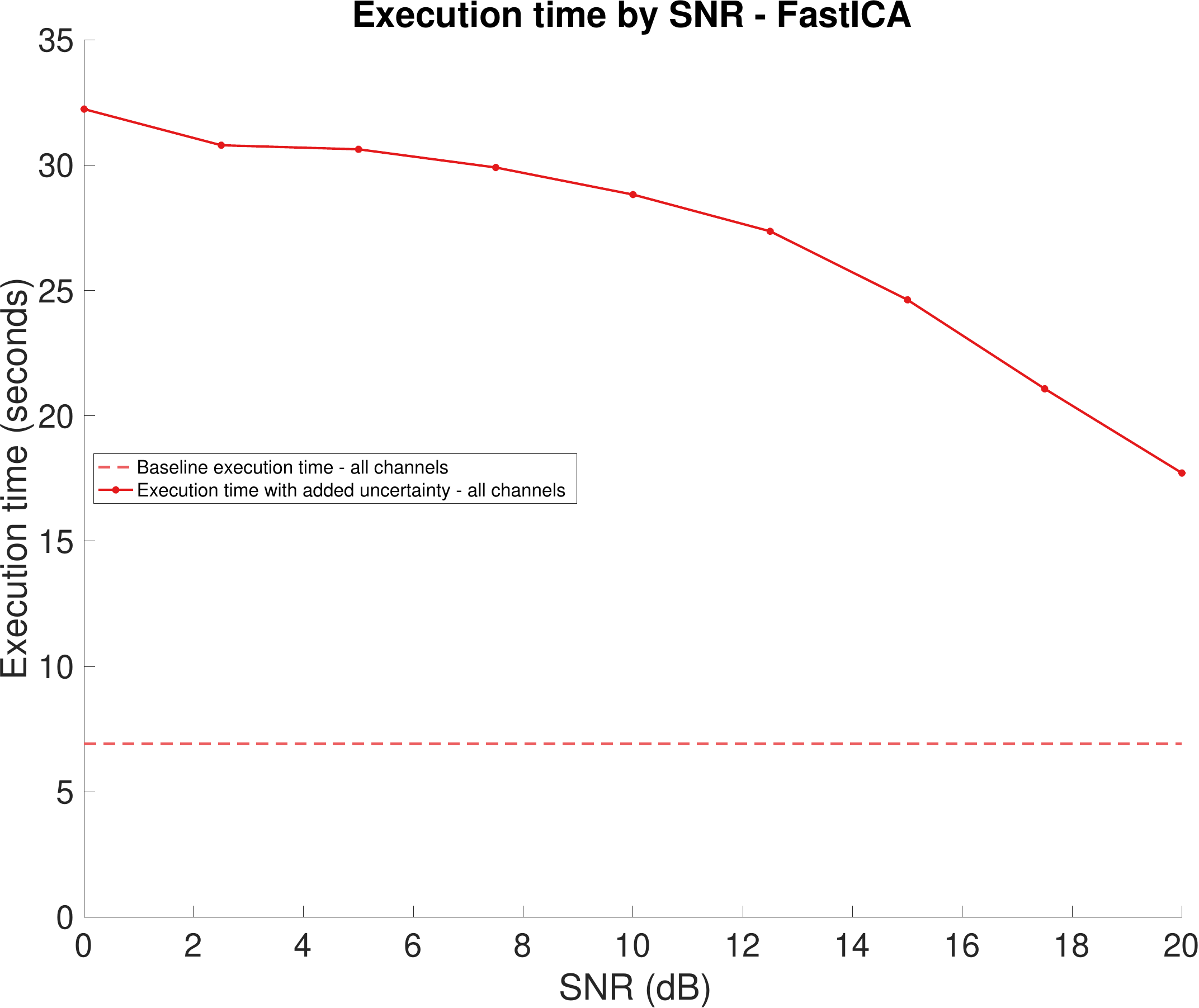}}
\hfil
\subfloat[]{\includegraphics[width=3in]{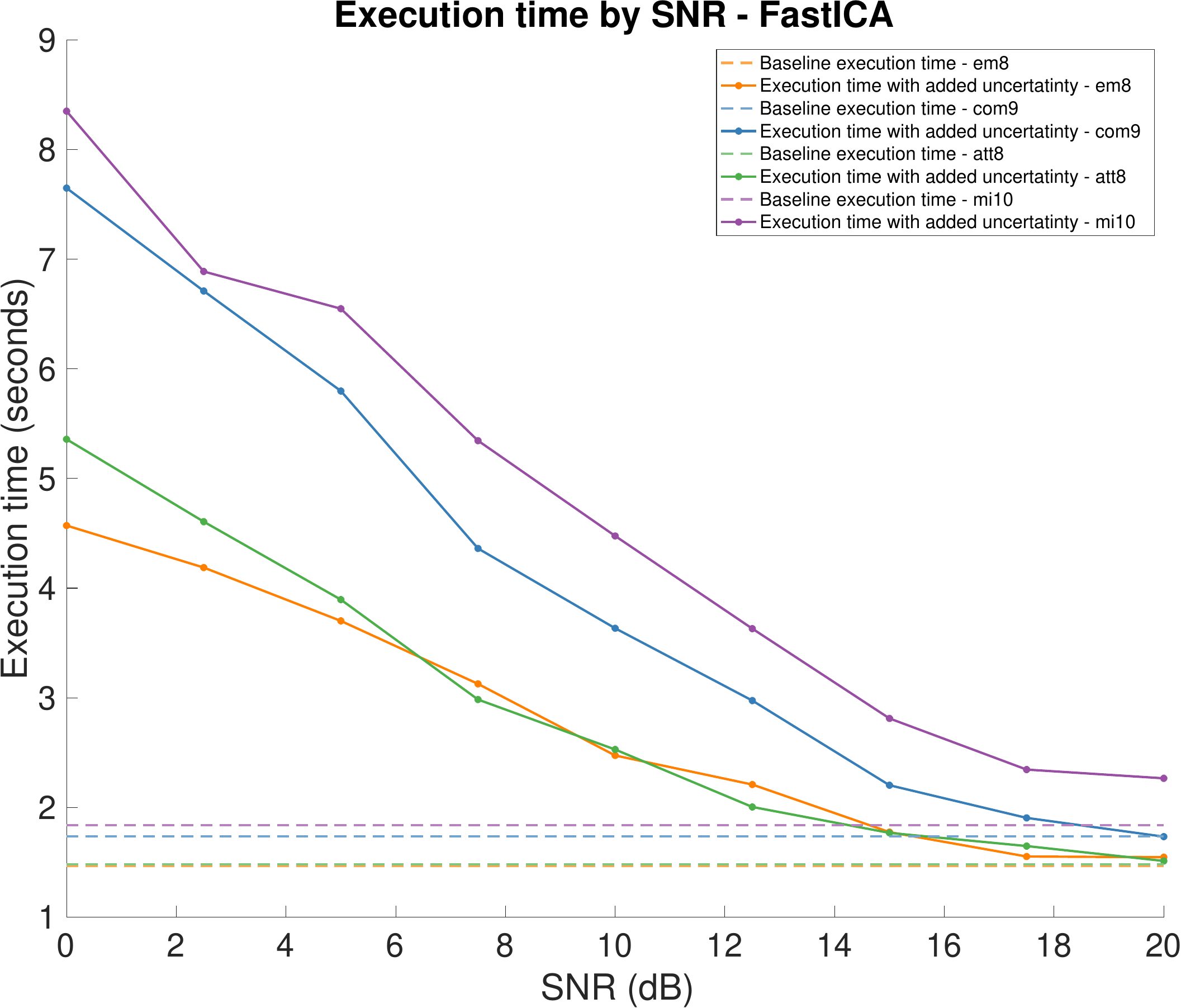}}
\vfil 
\subfloat[]{\includegraphics[width=3in]{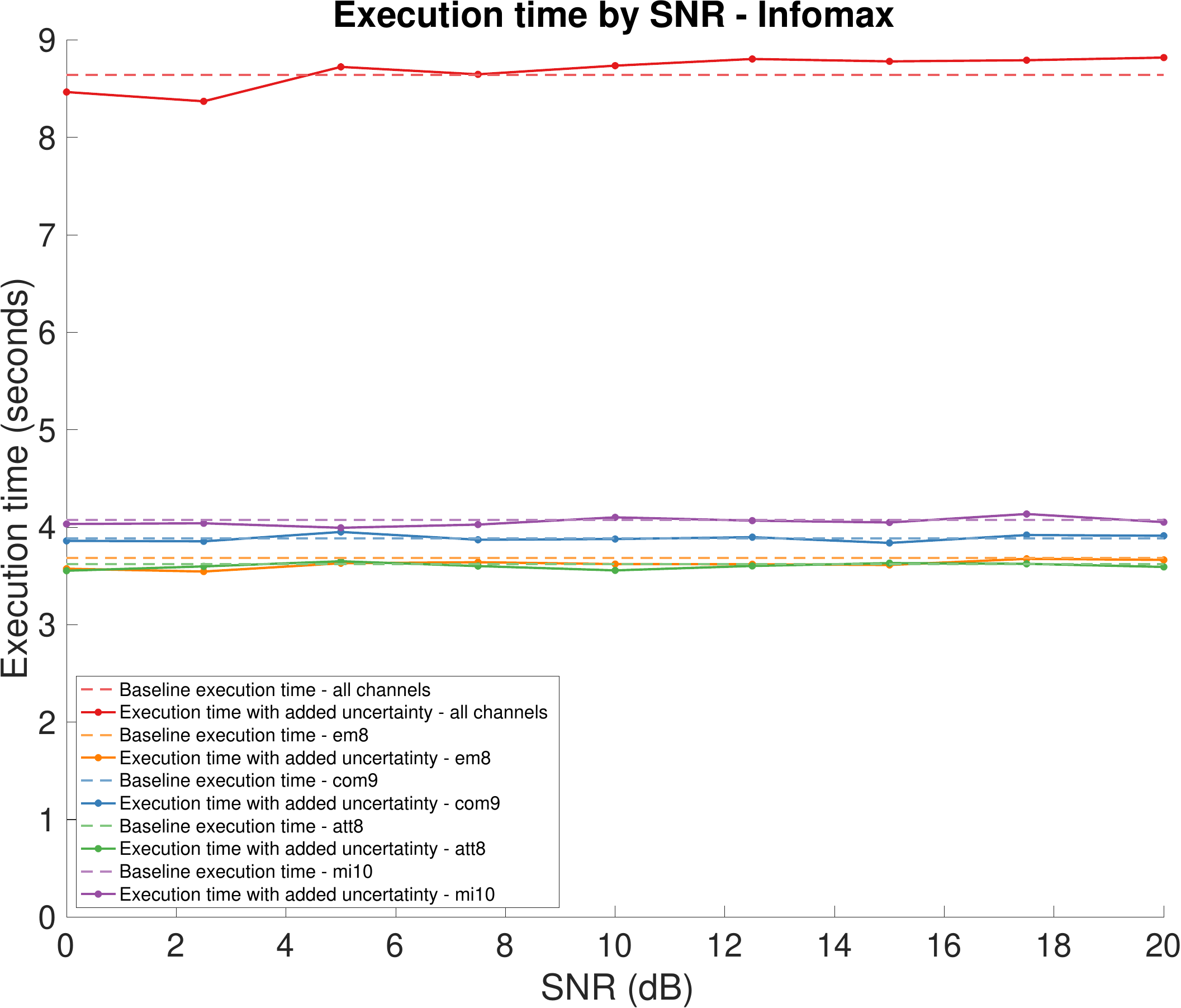}}
\caption{The average time for each Monte Carlo iteration to complete in each simulation.  We split the FastICA plots into two plots for more clarity - (a) shows the timing for simulations using all EEG channels, and (b) shows the timing for simulations using smaller subsets of electrodes. FastICA has a 50\%-85\% reduction in execution time over a 20\,dB increase in SNR.The baseline execution times for FastICA are always faster than the baseline execution times for Infomax for a given electrode configuration, however FastICA execution times are very susceptible to the level of noise.  
Conversely, Infomax execution times remain flat across all noise levels, demonstrating no dependence on SNR.}
\label{fig:Timing}
\end{figure*}

We also record the average execution time per re-execution for all of our simulations. 
Figure~\ref{fig:Timing} shows the average execution time of the 100 re-executions of the simulations.
FastICA has faster baseline execution times than Infomax for any given electrode configuration. 
However, FastICA execution times are dependent on the simulated SNR.
In the case where all electrode channels are used, FastICA's execution time is over
twice as slow as the baseline case even when the SNR is $20$\,dB.
For electrode configurations that include fewer channels, FastICA's execution time does approach the baseline time for higher SNR. 
Observing all of the electrode configurations together, we note that FastICA's execution time reduces by 50\%-85\% over an SNR increase of 20\,dB.
The execution time of Infomax is not dependent on the noise level 
and instead remains flat across all noise levels for all electrode configurations.
In essence, FastICA's execution time is sensitive to the amount of uncertainty present in the measurement, 
whereas Infomax's execution time is only a function of how many channels are being used.

The root cause of this difference in performance stems from how the two algorithms find each individual component. 
FastICA is a fixed-point algorithm that finds one independent component at a time. 
Therefore, any factor that has an impact on the convergence speed of the algorithm is amplified by the number of independent components being found.
In contrast, Infomax finds each component in parallel, so any factors that impact convergence speed will only have an effect one time, rather than being multiplied for each component. 

\section{Conclusion}
\label{sec:Conclusion}
We experimentally determined that the measurement uncertainty of the ADS 1299 follows a Gaussian distribution. 
We then ran several simulations of both FastICA and Infomax ICA using synthetically generated EEG data and adding measurement uncertainty modeled as additive Gaussian noise.
FastICA and Infomax show similar correlation performance for given SNR values and electrode configurations.
As expected, higher amounts of measurement uncertainty lead to lower correlation between the most likely eyeblink component and the reference eyeblink component.

For most of the electrode configurations we studied, an SNR of around $15$\,dB corresponds to an approximate $5\%$ performance degradation 
when compared to a baseline case where no measurement uncertainty is present.
This means that if the RMS amplitude of the biological signal is at least $5.6$ times higher than the RMS amplitude of the measurement uncertainty, 
there will likely be less than a $5\%$ difference in expected performance when compared to a scenario where there is no measurement uncertainty.
The biggest performance difference between FastICA and Infomax is in the execution times between the two algorithms.
FastICA's execution time is highly sensitive to measurement uncertainty, whereas Infomax's execution time is unaffected by the presence of measurement uncertainty
and instead is generally a function of the number of channels included in the analysis.
For SNRs greater than $10$\,dB FastICA consistently outperforms Infomax when working with a subset of the channels (i.e., em8, com9, att8, or mi10).

\bibliographystyle{IEEEtran}
\bibliography{IEEEabrv,references}

\end{document}